\documentclass[11pt]{article}
\usepackage{times,amscd, amssymb}
\usepackage{latexsym}
\usepackage{graphicx}
\usepackage{amsfonts}
\usepackage{amsmath}
\usepackage{verbatim}
\usepackage[english]{babel}
\usepackage[authoryear]{natbib}
\usepackage{xcolor} 
\usepackage{soul}
\RequirePackage[colorlinks,citecolor=blue,urlcolor=blue]{hyperref}
\usepackage[top=1.05in,left=1in,bottom=1in,right=1in]{geometry}  

\usepackage[switch]{lineno}

\newtheorem{defn}{Result}

\def\bSig\mathbf{\Sigma}

\newcommand{\Wb}{\overline}

\def\*#1{\boldsymbol{#1}} 
\def\##1{\mathbf{#1}} 

\title{Novel models of trait evolution via an expansion of Lande’s fitness function: The Ornstein-Uhlenbeck process meets the Little Prince’s boa}

\author
{Jos\'e M. Ponciano\footnote{
Department of Biology, University of Florida, Gainesville, FL. 32611.}, C. Godineau\footnote{
Department of Biology, University of Florida, Gainesville, FL. 32611.}, L. Jiménez\footnote{
Centro de Investigación e Innovación para el Cambio Climático, Facultad de Ciencias, Universidad Santo Tomás, Chile.}, N. Kortessis\footnote{Department of Biology, Wake Forest University, Winston-Salem, NC. 27109}, R. Zenil-Ferguson\footnote{Department of Biology, University of Kentucky, Lexington, KY. 40508},\\
and Robert D. Holt\footnote{Department of Biology, University of Florida, Gainesville, FL. 32611.}}
\date{}

\begin{document}
\maketitle

\begin{abstract}
Adaptive topographies form the foundation for much of our understanding of evolutionary change. Lande's 1976 influential paper on the adaptive topography of phenotypes demonstrated how the concept is inherent in both phenotypic and genetic models of evolution, and how the concept can be used to test evolutionary hypotheses given data. Here, we revisit and generalize Lande’s original derivation of an equation analogous to Wright’s genotypic adaptive topography to the case of two fitness components. A move to two fitness components yields novel predictions about the shape and mechanistic underpinnings of the adaptive topography. The optimum of this updated fitness function is a weighted average of the optima of the two fitness components, with weights given by the relative strengths of stabilizing selection on each component. Temporal or spatial heterogeneity in the strengths of selection for each fitness component create novel shapes (asymmetry, bi-modality, or lack thereof) of the overall fitness function, a possibility demonstrated with a case-study from the published literature. Finally, when combined with Lande's approach to generate an Ornstein-Uhlenbeck (OU) model for the evolution of the average phenotype, our fitness formulation leads to a previously unrecognized family of stochastic differential equation models of trait evolution. These results provide mechanistic justification for non-Gaussian fitness functions (often observed in natural systems), provide a path for testing alternative models generating non-Gaussian fitness functions, and pave the way for future study of the interplay of ecological and evolutionary dynamics, such as in the study of evolutionary rescue.

\end{abstract}


\newpage
\section*{Non-specialist summary (119 words):}
The fitness landscape --- a mapping from phenotypic traits to Darwinian fitness--- is classically treated as a static surface with fixed widths around adaptive peaks. Yet natural environments fluctuate, and so do the relative strengths of selection acting on survival and reproduction. We show that when these component-specific selection strengths vary, adaptive peaks move even when the underlying selective targets remain fixed: the effective optimum is a weighted average of reproduction and survival optima, and the weights are their respective relative strengths of stabilizing selection. Understanding spatio-temporal fitness landscape changes and why they happen is a fundamental open problem in evolutionary biology.  Our results have broad implications for macroevolutionary inference, comparative phylogenetic methods, and the theory of evolutionary rescue.

\newpage
\noindent \textit{Left to our own devices we are not very good at picking out patterns from a sea of noisy data. To put it in another way, we are all too good at picking out non-existent patterns that happen to suit our purposes.}
\begin{flushright}-Efron and Tibshirani 1993.
\end{flushright}

\noindent \textit{Nature laughs at the difficulties of integration.}
\begin{flushright}-Pierre Simon de Laplace.
\end{flushright}

\section*{Introduction}
\label{s:intro}

Evolutionary biology aims to understand how traits change and lineages diversify. Modeling the link between the shape of the fitness landscape and the rate and direction of evolutionary change results in a better understanding of trait evolution and diversification. Importantly, that link drives work on \textit{e.g.} the evolvability of antibiotic resistance \citep{Jordt:2020}, the evolution of extinction risk in conservation biology \citep{Kortessis:2023}, and the diversification of tropical birds \citep{Gomez:2020}. These examples depend on a clear description of how the slope and curvature of the fitness surface influence evolutionary dynamics. In this paper, we revisit the foundational work of \cite{Lande:1976} characterizing the shape of fitness landscapes defined over phenotypes and show that incorporating aspects of heterogeneity (in the environment, or among fitness components) can lead to non-Gaussian fitness surfaces that ultimately govern evolutionary trait change. Our approach generates new models for examining the evolution of traits in a non-constant fitness landscape. 

\cite{Wright:1932} first tied allele frequency change to the way mean fitness responds to shifts in allele frequency. Forty four years later, \cite{Lande:1976} built the phenotypic analogue. He linked selective change in a population’s mean phenotype to the derivative of mean fitness with respect to mean phenotype. By connecting the breeder’s equation \citep{Falconer:1960} to stabilizing selection, Lande showed that selection acts through the local geometry of the fitness surface. A narrow peak creates a strong pull toward an optimum. This insight led naturally to models that resemble Ornstein–Uhlenbeck (OU) dynamics, in which curvature controls the strength of evolutionary attraction. OU models and their extensions \citep{Hansen:1997,Butler:2004,Uyeda:2011,Uyeda:2014,Pennell:2014,Uyeda:2015,Boucher:2018,Uyeda:2018,Blomberg:2020,Week:2021,Boyko:2023} now sustain much of macroevolutionary theory. Because these evolutionary models emerge directly from assumptions about the geometry of the fitness surface, understanding the processes that shape adaptive landscapes is central to understanding evolutionary dynamics themselves.

Evolutionary models have traditionally accommodated environmental heterogeneity by allowing the optimal phenotype favored by selection to shift across space or through time while keeping the shape of the fitness surface fixed \citep{kopp:2014}. Less attention has been given to the possibility that environmental variation may instead alter the width of fitness functions, thereby changing the strength of stabilizing selection itself. Yet, empirical studies routinely find variation in the strength of selection. Selection can tighten or weaken along climatic, resource, or predator gradients \citep{benkman:1999,benkman:2003,etterson:2004}, and field studies have reported inter-annual shifts in both the strength and form of selection \citep{reimchen:2002,carlson:2007,brown:2013,vanBuskirk:2021}. Broad syntheses confirm that nonlinear selection varies widely across taxa, environments, and years \citep{Kingsolver:2001,Siepielski:2009}.

At the same time, natural selection acts through multiple fitness components — such as survival, growth, and reproduction — that often favor different trait values and impose different selection curvatures \citep{arnold:1984a, arnold:1984b, kingsolver:2007, Kingsolver:2011}. Because organisms succeed only when they pass through these components, total fitness arises as the product of component‑specific fitness functions implying a structure that can reshape the composite fitness surface in ways no single component predicts \citep{shaw:2010}. Previous work has shown that this structure can generate complex fitness surfaces and complicate statistical inference \citep{mitchell:1987,shaw:2008,shaw:2010}.

When fitness arises from multiple components, variation in fitness-function width may be particularly important because it can influence how strongly each component contributes to the resulting fitness landscape. Consequently, variation in fitness-function width has the potential to affect not only the strength of selection, but also the location, curvature, and overall shape of the composite adaptive landscape. Changes in component-specific fitness-function widths may shift effective adaptive optima even when the underlying component-specific optima remain unchanged, providing a potential mechanism for moving between adaptive peaks without changes in the selective targets themselves. Despite widespread empirical evidence for variation in selection strength, and the central role of multiple fitness components in determining lifetime fitness, the consequences of component-specific fitness-function widths for overall adaptive-landscape geometry remain poorly understood.

These considerations led us to examine how component-specific selection and variation in selection width across space and time combine to shape the composite fitness surface. We first review core elements of the evolution of a single quantitative trait \cite{Lande:1976,lynch1998genetics,arnold2023evolutionary}. We develop a new model that captures the consequences of component-specific selection. Initially, and to a first approximation, each fitness component has a Gaussian shape but may differ in optimum and width. We first show that the product of Gaussian fitness components generates a Gaussian composite landscape whose optimum is determined by a width-weighted combination of component-specific optima. We then allow for environmentally-driven random shifts in these widths across space or time and specify a general hierarchical model for the composite fitness function. Under this formulation, heterogeneity in stabilizing-selection strength generates effective fitness landscapes that are generally non-Gaussian despite each fitness component being Gaussian. We use this framework to determine when spatial variation produces non-quadratic composite landscapes, and when temporal variation shifts the effective optimum by changing the relative weight of component optima.  We then insert these new landscapes into Lande's original model derivation of the adaptive topography and his continuous stochastic process of phenotypic evolution. In so doing, a new family of alternative diffusion process models of trait evolution emerges. This new family provides a novel gateway into linking different heterogeneities in the strength of selection with different patterns of phenotypic macroevolution.  We discuss unresolved statistical parameter identifiability challenges while developing a proof-of-concept data analysis example using the well-known dataset of \cite{etterson:2004}. Finally, we discuss the implications of our findings in light of current paradigms in phenotypic diversity maintenance, evolutionary rescue, and mutation load.


\section*{Wright and Lande's adaptive topographies}
Sewall Wright's genotypic adaptive topography model \citep{Wright:1932,Wright:1937, Wright:1942,Wright:1982, Wright:1988} states that $\Delta p$, the change in the frequency of an allele $p$, depends on the mean fitness $\overline{W}$ according to the relation 
$$
\Delta p = \frac{p(1-p)}{2\overline{W}}\frac{\partial \overline{W}}{\partial p} = \frac{p(1-p)}{2}\frac{\partial {\rm ln}\,\overline{W}}{\partial p},
$$

\noindent Lande's adaptive topography for phenotypes, also obtained under the assumption of discrete generations, recognizes that the response to selection described by the breeder's equation \citep{Falconer:1960} can be spliced with a quantitative description of the force of selection as described by the change in average fitness $\Wb{W}$ in a generation. 

We make $p(z,t)$ the probability density function (pdf) of a phenotypic trait $z$ in generation $t$ of the population of interest before selection, and $W(z)$ the ``fitness function'' describing how the phenotypic character influences the absolute fitness of an organism. As well, let 
\begin{equation}
p_{W}(z,t) = \frac{W(z)p(z,t)}{\int W(z)p(z,t) dz} = \frac{W(z)}{\Wb{W}} p(z,t)
\end{equation}
\noindent be the pdf of the distribution of the phenotypic character after selection in the population, with ${\Wb{W}}$ being the population mean fitness. Using this distribution, one can compute the expected value of the phenotype after selection $\Wb{z}_{W}(t)$ as
\begin{equation}\label{intaphas}
\begin{array}{lll}
\Wb{z}_{W}(t) &=& {\rm E}_{p_{W}}[Z_{w}] = \int zp_{W}(z,t)dz = \frac{\int z W(z)p(z,t) dz}{\int W(z)p(z,t) dz}\\
 & & \\
 &=& \frac{1}{\Wb{W}}\int z W(z)p(z,t) dz.
\end{array}
\end{equation}
\noindent Then, using a normal probability density function (pdf) for $p(z,t)$ with mean $\Wb{z}(t)$ and variance $\sigma^2$, it follows  that the change in the average fitness is given by 
\begin{equation}\label{changefit}
\frac{\partial \Wb{W}}{\partial \bar{z}(t)} = \frac{\Wb{W}}{\sigma^{2}}[\Wb{z}_{W}(t) - \Wb{z}(t)],
\end{equation}
\noindent where the difference $[\Wb{z}_{W}(t) - \Wb{z}(t)]$ is the selection differential (see derivation in Appendix \ref{ALandemath}).  This difference also appears in the breeder's equation describing the change in the average value of a phenotypic character in response to selection in one generation, $\Delta \Wb{z}(t) = [\Wb{z_{W}}(t) - \Wb{z}(t)]h^2,$ where $h^2$ is the proportion of genetic variance in the overall variance of a trait.  Solving for the selection differential in equation \ref{changefit}) and replacing the resulting expression into the breeders' equation allows one to rewrite the equation for the change in the average phenotype, $\Delta \bar{z}(t)$, as 
\begin{equation}\label{LAT}
\Delta \Wb{z}(t) = \frac{\sigma^{2}h^{2}}{\Wb{W}}\frac{\partial \Wb{W}}{\partial \bar{z}(t)} = \sigma^{2}h^{2} \frac{\partial {\rm ln}\, \Wb{W}}{\partial \bar{z}(t)},
\end{equation}
\noindent which is parallel to Wright's 1937 equation for gene frequencies, where the heritable allelic variance, $p(1-p)/2$, is the allelic analogue to heritable phenotypic variance, $\sigma^2h^2$, but for phenotypes. Uniting these two perspectives is the combination of heritable variation in the population and the action of selection on that variation, i.e., the adaptive topography. For reflections on the utility and limitations of adaptive landscapes see \cite{svensson2012adaptive}.

The adaptive topography is analogous to the contours of the individual fitness function, $W(z)$, but applied at the population level rather than at the level of individual phenotypes. It represents how the mean phenotype “moves” on a version of the original fitness surface smoothed by the phenotypic variance. In what follows, we explore first how this adaptive topography may change under different conceptualizations of the fitness function.  From there, we follow in Lande's footsteps to propose a class of stochastic models of trait evolution using these different conceptualizations of the fitness function, a process that yields a novel family of stochastic models of trait evolution. 

\section*{Fitness components and the adaptive topography of phenotypes}
The fitness model \cite{Lande:1976,Lande:1979} used to derive the OU model of phenotypic evolution assumes Gaussian stabilizing selection with an optimum phenotype set at zero. The choice of a Gaussian is, in part, an appeal to mathematical convenience, but is also a sensible choice in the absence of further information, assuming the central limit theorem to represent the hypothesis that fitness results from the effects of many independent processes of small effect. Lande further justified the choice as follows: “because any smooth fitness function can be closely approximated by a Gaussian function in the vicinity of the optimum” \citep[][p. 322]{Lande:1976}. Essentially, Lande argues that it is enough to look at how fitness varies in the vicinity of the optimum, if it exists, and that one can set this optimum to zero such that scale and translation of the measure of the phenotypic character (which become nuisance variables) can be safely ignored. Here, we investigate the consequences of ignoring allowing for different optima for different fitness components, as well as deviations from a smooth, symmetrical curvature near an overall fitness optimum.

Among all the functions describing how fitness varies as a function of a trait, the ones whose shape arises naturally as a function of basic biological principles allow experimentalists to connect evolutionary and ecological processes with observed patterns in nature. Fundamentally, determining the shape of fitness functions in nature is an empirical issue, yet biologically grounded models should be useful tools for estimating such shapes and making predictions about evolutionary change. Our goal is to explore how non-Gaussian models of fitness may arise and their evolutionary as well as inferential consequences. Doing so is a crucial step in developing statistical models for use in inferring evolutionary processes. This is even more important in light of questions about the use of complicated mathematical functions that are justified on a match between observed patterns in nature. Some have argued against such models because they may, in the end, not bring as much understanding and insight as simpler models which seemingly lack relevant biological detail \citep[see for instance][]{Berryman:1985, Strong:1999}. 

\subsection*{Constant selection strength acting on different fitness components}
\subsubsection*{The adaptive topography through two-stages life cycle}
One way to construct a Gaussian fitness function is to suppose that for a given organism with character $z$ (\textit{e.g.} body size), there are a large number $B$ of individuals with each individual $i$ having a chance $F_i(z)$ of successfully reproducing during their lifetimes (and chance $F_i^c(z) = 1 - F_i(z)$ of failing to reproduce.  We assume that all individuals that do reproduce have the same expected number of offspring $\beta$, wich we will come back to). The fitness of a character then reflects the average fitness across all individuals with said character. As such, the average chance of reproducing across individuals is $B^{-1}\sum_{i=1}^BF_i(z)= 1 - B^{-1}\sum_{i=1}^BF_i^c(z)$. Assuming independence among individuals in their realized reproduction, the average chance of reproduction for an individual with trait $z$ is then 
$$
\Wb{F(z)} \approx \left(1-\frac{1}{B}\sum\limits_{i=1}^{B}F_{i}^{c}(z)\right)^{B} \approx {\rm e}^{-\sum\limits_{i=1}^{B}F_{i}^{c}(z)},
$$
\noindent where the second approximation holds for sufficiently large $B$. Now, consider $h(z) = \frac{1}{B}\sum\limits_{i=1}^{B}F_{i}^{c}(z)$, the average chance of reproductive failure for a character $z$. The optimal trait value $z_{F}^\star$ (if it exists) with respect to reproduction is then defined as the value of $z$ that minimizes $h(z)$. That is, $h'(z_{F}^{\star}) =0$ and $h''(z_{F}^{\star}) >0$. Accordingly, $h(z_{F}^{\star})\geq0$ is the smallest value that this sum attains. If $h(z_{F}^{\star})$ is exactly zero, then all individuals with the value $z=z_{F}^{\star}$  reproduce. A Taylor Series approximation around this optimum yields 
\begin{equation}
\begin{split}
h(z) & = h(z_{F}^{\star}) + h'(z_{F}^{\star})(z-z^*_F) + \frac{h''(z_{F}^{\star})}{2}(z-z_{F}^{\star})^2 + o((z - z_F^{\star})^2) \\
& = h(z_{F}^{\star}) + \frac{(z-z_{F}^{\star})^2}{2\frac{1}{\delta}} + o((z - z_F^{\star})^2),
\end{split}
\end{equation}
\noindent where the linear term is dropped in the final line because $h^\prime(z_F^\star) = 0$ and where we have introduced $\delta=h''(z_{F}^{\star})$, the curvature of reproductive success around the optimum. Substituting this equation into the expression for the average chance of reproduction, we have exactly a Gaussian kernel
\begin{equation}\label{gaussreprod}
\Wb{F(z)}\approx {\rm e}^{-h(z_F^\star)}{\rm e}^{-\frac{(z-z_{F}^{\star})^2}{2\frac{1}{\delta}}},
\end{equation}
where $\mathrm{e}^{-h(z^*_F)}$ is a constant related to maximum reproduction that does not affect the shape of the fitness function, and so only influences absolute fitness, but not relative fitness. 

By direct analogy with maximum likelihood theory, the quantity $\delta=h''(z_{F}^{\star})$ (the curvature of $h(z)$) contains information about how much the probability of not reproducing is changed for different values of  $z$ near the trait optimum. We present it in the expression above as a reciprocal, $\frac{1}{\delta}$, as it plays the role of the variance in a Gaussian kernel. However, just as in Bayesian statistics, where for the sake of analytical tractability it is usually accustomed to work with the inverse of the variance parameters of Gaussian kernels (the so-called ``precision parameters''), we emphasize the parameter $\delta$ because it is a measure of the strength of selection. All else equal, the strength of selection near the optimum increases for larger values of $\delta$.  

The development above is a suitable description of a key aspect of fitness (the probability of successful reproduction) for semelparous species with nonoverlapping generations and where the character $z$ affects the chance of reproducing. It shows that under the assumption of a single optimum trait, one may arrive at a formulation of the stabilizing-selection Gaussian fitness model (see \citealp[p 322, eq. 13]{Lande:1976}, and \citealp[eq. 2.3]{Turelli:1984}).  But the argument applies more generally, as Lande suggested, when the fitness function has a single optimum and selection is not too strong. 

\subsection*{Dual fitness function}
One can extend this model by assuming that individual survival up to reproduction is also influenced by trait $z$. Using an argument analogous to the reproduction argument above, one may approximate the average probability of surviving $\Wb{S(z)}$ as a function of the trait $z$, its optimum $z_{S}^{\star}$ and its associated strength of stabilizing selection $1/\gamma$ as             
\begin{equation}\label{gausssurv}
\Wb{S(z)} \approx {\rm e}^{-h_S(z_S^\star)}{\rm e}^{-\frac{(z-z_{S}^{\star})^2}{2\frac{1}{\gamma}}}.
\end{equation}
\noindent Using equations (\ref{gaussreprod}) and (\ref{gausssurv}) and Result 1 in Appendix \ref{AGausskernel}, a general fitness function can then be written as 
\begin{align}\label{newfitness}
W(z) &= \text{\small{offspring per reproductive bout}} \times {\rm P}[\text{reproducing}|\text{survival}] \times {\rm P}[\text{survival}],\nonumber\\
&=  \beta\times {\rm e}^{-\frac{(z-z_{F}^{\star})^2}{2\frac{1}{\delta}}}{\rm e}^{-\frac{(z-z_{S}^{\star})^2}{2\frac{1}{\gamma}}} = \beta {\rm e}^{-\frac{(z-z^\star)^{2}}{2\omega^{2}}}{\rm e}^{-(z_{F}^{\star}-z_{S}^{\star})^{2}/2\left(\frac{1}{\delta} + \frac{1}{\gamma}\right)},
\end{align}
\noindent where $\beta$ is a constant, and
$$
z^\star = \frac{\frac{1}{\gamma}z_{F}^{\star} + \frac{1}{\delta}z_{S}^{\star}}{\frac{1}{\delta} + \frac{1}{\gamma}} \quad\text{and}\quad \omega^{2} = \frac{1/(\delta\gamma)}{\frac{1}{\delta} + \frac{1}{\gamma}}=\frac{1}{\delta + \gamma}
$$
\noindent represent the optimum trait and strength of selection integrating both survival and fecundity selection, respectively. Thus, when selection acts \textit{via} the probabilities of reproduction and survival, the resulting fitness function is also a Gaussian function. Remarkably, the total strength of selection on $z$ is simply $1/\omega^2 = \delta + \gamma$, i.e., the sum of selection strengths on survival and fecundity separately. Moreover, the optimum trait integrates both sources of selection and can be interpreted as a weighted average of the optima for each component separately. One can see this because $z^*$ can be written as 
\begin{equation}\label{thetaweights}
    z^* = \mathcal{P}z^\star_F + (1-\mathcal{P})z^\star_S,
\end{equation}
\noindent where 
$$\mathcal{P} = \frac{\gamma^{-1}}{(\delta^{-1} + \gamma^{-1})} = \frac{\delta}{\gamma + \delta}$$ 
\noindent is the relative \textit{contribution} of fecundity to the total strength of selection, and so 
$$1-\mathcal{P} = \frac{\gamma}{\gamma + \delta}$$ 
\noindent is the \textit{contribution} of survival to the total strength of selection). 

Finally, note that an explicit consideration of the two fitness components reveals the fact that the maximal fitness is depressed by the quantity ${\rm e}^{-(z_{F}^{\star}-z_{S}^{\star})^{2}/2\left(\frac{1}{\delta} + \frac{1}{\gamma}\right)}$, which is a number between 0 and 1. In particular, discordance between the locations of the reproductive and survival optima, given by $(z_{F}^{\star}-z_{S}^{\star})^{2}$,  reduce the total fitness function, with the reduction scaled by the combination of both strengths of selection $\frac{1}{\delta} + \frac{1}{\gamma}$. 

These results illustrate that Lande's derivation of a quadratic fitness function is robust, even under stabilizing selection on two separate fitness components. The novelty of this analysis is that it reveals a mechanistic understanding of the resulting trait optimum as coming from selection on multiple fitness components (i.e., a weighted average of component optima). In particular, the overall trait optimum is shaped by the relative strengths of selection on each component. As we will show below, the fact that the trait optimum may itself be determined by the strength of selection has important implications for statistical estimation of trait optima and strength of selection parameters of phenotypic characters, and for their interpretation. We note that although we have have assumed that fecundity is constant given survival and reproduction, another avenue for expanding these ideas is to set $\beta$ equal to a function of the trait, $\beta(z)$. 


\subsection*{Heterogeneity in selection strength acting on different fitness components: the dual fitness function revisited}
\subsubsection*{Temporal heterogeneity}\label{section:vertical heterogeneity}
Heterogeneity in selection strength can be introduced in more than one way. For instance, one could vary $\delta$ and $\gamma$ randomly \textit{over time} in \ref{newfitness}, from generation to generation, a form of temporally heterogeneous selection. This is variation in selection that happens from one generation to the next, not during multiple episodes of selection in a generation. Operationally, to generate this type of heterogeneity, one can let the selection strengths $\delta$ and $\gamma$ vary randomly and independently \textit{over time} according to two different gamma distributions with the same rate.  That is, we let $\delta\sim\mathrm{Gamma}(\alpha_F,b)$ and $\gamma\sim\mathrm{Gamma}(\alpha_S,b)$ with the same rate parameter $b$. Then, $\mathcal{P}$ in the optimum $\mathcal{P}z_{F}^{\star} + (1-\mathcal{P})z_{S}^{\star}$ may be assumed to be Beta distributed. The total strength of stabilizing selection $\delta+\gamma$ is then Gamma distributed with shape parameter $\alpha_F + \alpha_S$ and rate $b$. Below we will make explicit this idea, but before doing so, we expand on another type of heterogeneity in selection strength representing spatial heterogeneity.

\subsubsection*{Spatial heterogeneity: random effects}
The above developments represent just the tip of the iceberg of what is possible to ask, once one is open to a dual fitness function. If we were to ask ``which biological processes or mechanisms could give rise to asymmetric fitness functions?'', an evolutionary ecologist would readily hypothesize that some form of heterogeneity could be at play.  Let us hypothesize, for instance, that heterogeneity in the strength of stabilizing selection for reproduction, survival, or both is at play. Going back to our Gaussian kernels for reproduction and survival (equations \ref{gaussreprod} and \ref{gausssurv}, respectively), one could model these heterogeneities using, for example, Gamma distributions \textit{in lieu} of $\delta$ and $\gamma$ and integrate them as random effects. Biologically, such heterogeneity could represent spatial variation in the strength of selection across habitats or local environments. The resulting fitness function therefore describes the average fitness landscape experienced by a population distributed across heterogeneous environments as in the multiple niche model of \cite{Levene:1953}.

\begin{figure}[htbp]
\centering
\includegraphics[angle=0,width=0.99\textwidth]{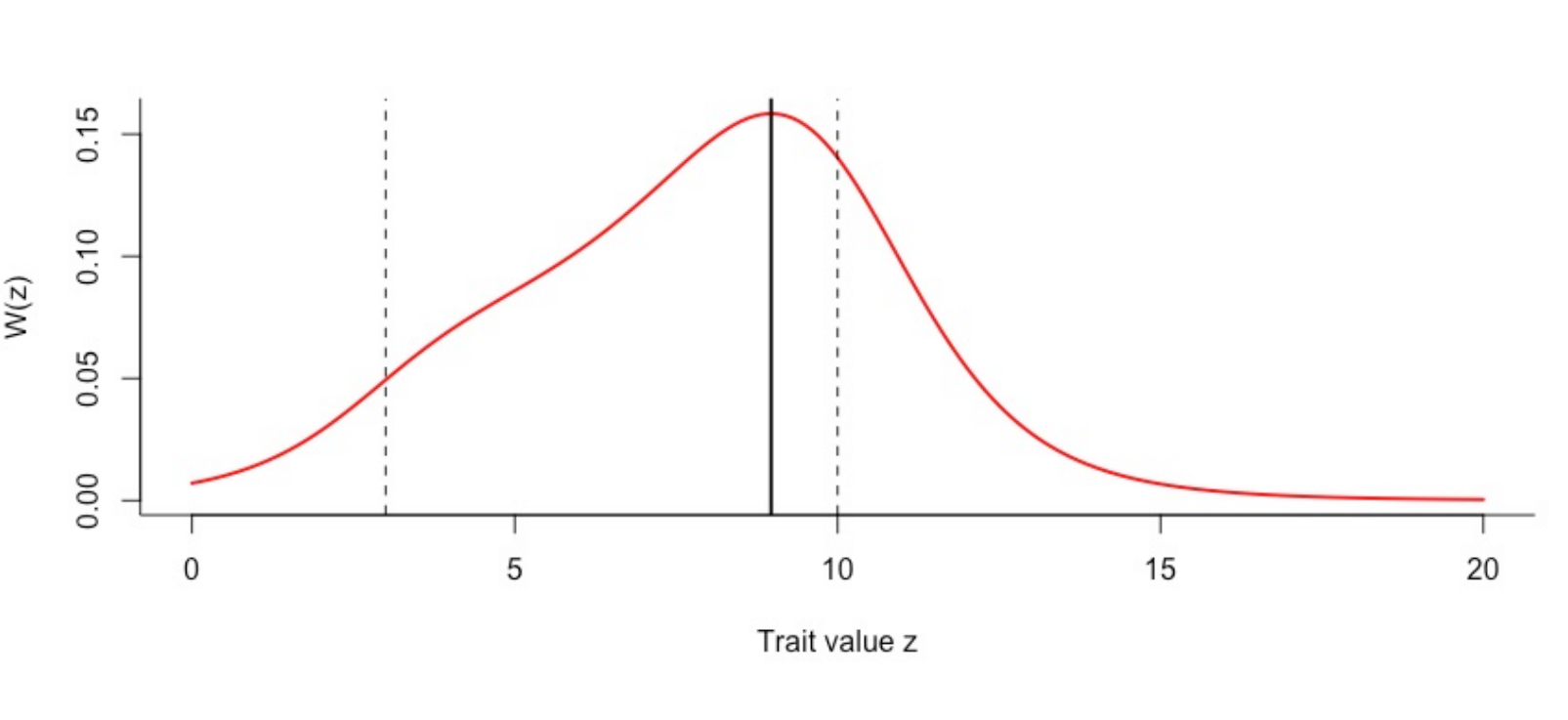}
\caption{A two-component fitness function where heterogeneity in the strength of selection for both reproduction and survival is assumed. Here, $a_{F}=1, b_{F}=4, a_{S}=2, b_{S}=7, z_{F}^{\star}=3, z_{S}^{\star}=10$.  The dotted vertical lines mark the locations of the optima for reproduction (at $z_{F}^{\star}=3$) and survival (at $z_{S}^{\star}=10$), respectively, whereas the solid vertical line denotes the joint optimum at $8.969$.}\label{Fig1}
\end{figure}

Let $\delta\sim\mathrm{Gamma}(a_F,b_F)$ and $\gamma\sim\mathrm{Gamma}(a_S,b_S)$, with pdfs \[ f_{\delta}(x)=\frac{b_F^{a_F}}{\Gamma(a_F)}x^{a_F-1}e^{-xb_F}, \qquad f_{\gamma}(x)=\frac{b_S^{a_S}}{\Gamma(a_S)}x^{a_S-1}e^{-xb_S}. \] Then, our fitness function transforms to
\begin{align}\label{heterofit}
W(z) &= \beta \left(\int{\rm e}^{-\frac{(z-z_{F}^{\star})^2}{2\frac{1}{\delta}}}f_{1}(\delta;a_{F},b_{F})d\delta\right) \left(\int{\rm e}^{-\frac{(z-z_{S}^{\star})^2}{2\frac{1}{\gamma}}}f_{2}(\gamma;a_{S},b_{S})d\gamma\right) \nonumber\\
\nonumber\\
&\propto \frac{1}{\left(1 + \frac{1}{2b_{F}}(z-z_{F}^{\star})^{2}\right)^{a_{F}}}\frac{1}{\left(1 + \frac{1}{2b_{S}}(z-z_{S}^{\star})^{2}\right)^{a_{S}}},
\end{align}
\noindent where each one of the large fractions in the equation above corresponds to the kernel of a generalized t-distribution. Note that we have used a Gamma distribution here, but many other heterogeneity distributions (models) are possible. When plotted, the product of these two non-central t-distribution kernels not only exhibits substantial asymmetry but, depending on the value of the Gamma distribution parameters, can lead to a hint of an incipient formation of a second mode, besides the largest mode (see Figure \ref{Fig1}). The largest mode, which is the joint optimum after selection in both reproduction and survival, has an exact, yet unwieldy analytical form that can be computed by taking the derivative of the logarithm of the function, setting it equal to 0 and solving for the value $z$; this joint optimum can be approximated as a weighted average.

\begin{figure}[htbp]
\centering
\includegraphics[angle=0,width=0.9\textwidth]{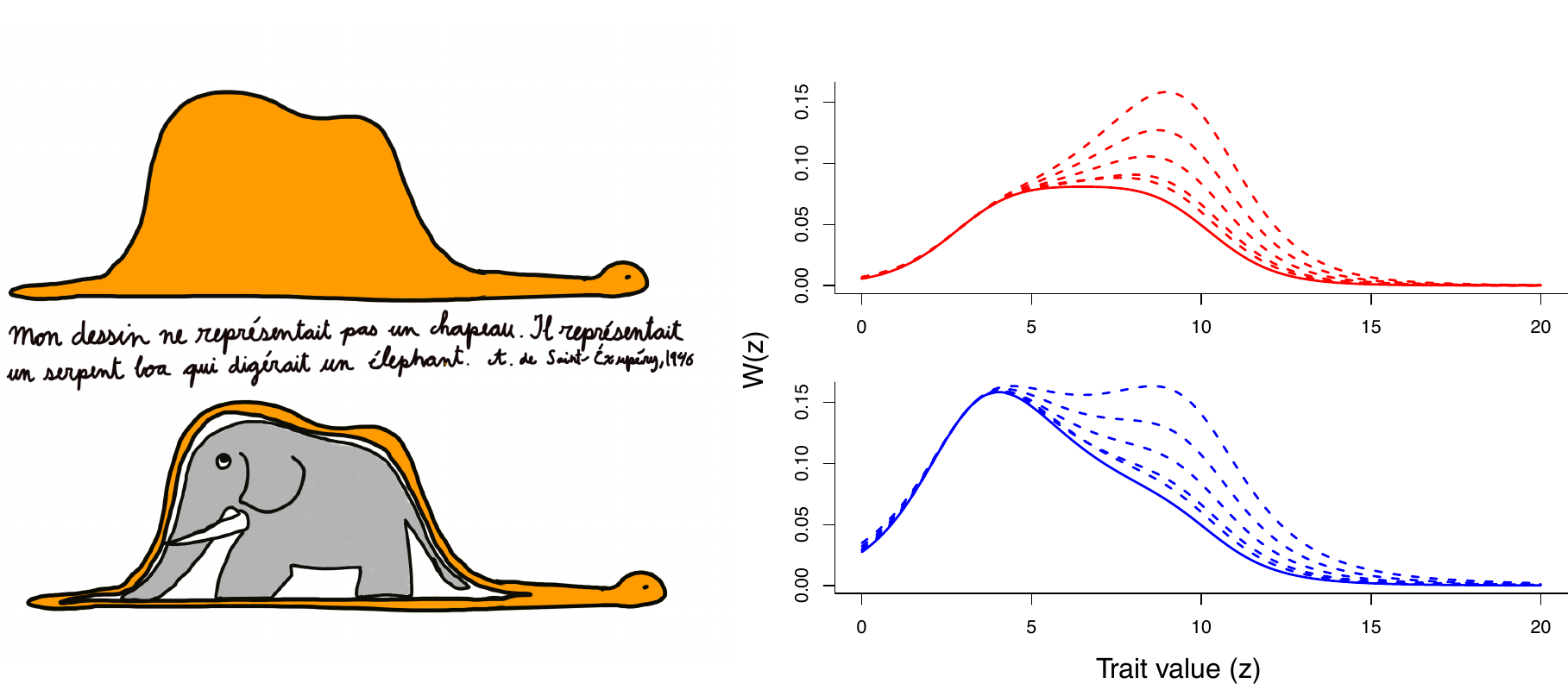}
\caption{Graphs on the right: Varying the relative sizes of the total amount of heterogeneity in the strengths of stabilizing selection for both reproduction (blue) and survival (red) leads to a wide array of fitness function forms (dashed lines) wherein  ``the elephant inside the Little Prince's boa'' on the left can go from being in a pronounced standing position so to speak, to a downright ``lying down'' position.  Left:  drawing reproduced freehand by J.M.P. from his 1979 copy of ``Le Petit Prince'', A. de Saint-\'Exup\'ery, 1946, wherein a boa constrictor was depicted as having gulped an elephant.}\label{Fig2}
\end{figure}

The comparative amount of heterogeneity in the strength of selection in reproduction and survival dramatically alters the shape of the fitness function.  In Figure \ref{Fig1}, the variance of the strength of selection in reproduction is $V(\delta)=a_{F}/b_{F}^{2}\approx 0.08533$ whereas the variance of the strength of selection in survival is $V(\gamma)=a_{S}/b_{S}^{2} \approx 0.04082$. The incipient mode on the left side, close to $z=5$, suggests that varying the relative sizes of these heterogeneities should lead to a wide array of fitness function forms.  Indeed, Figure \ref{Fig2} illustrates that when the relative sizes (importance) of these heterogeneities are varied, a wide diversity of fitness function shapes emerges. Moreover, bimodality in the fitness function emerges.

\subsubsection*{Illustration of the spatial heterogeneity model}
\begin{figure}[htbp]
\centering
\includegraphics[angle=0,width=0.9\textwidth]{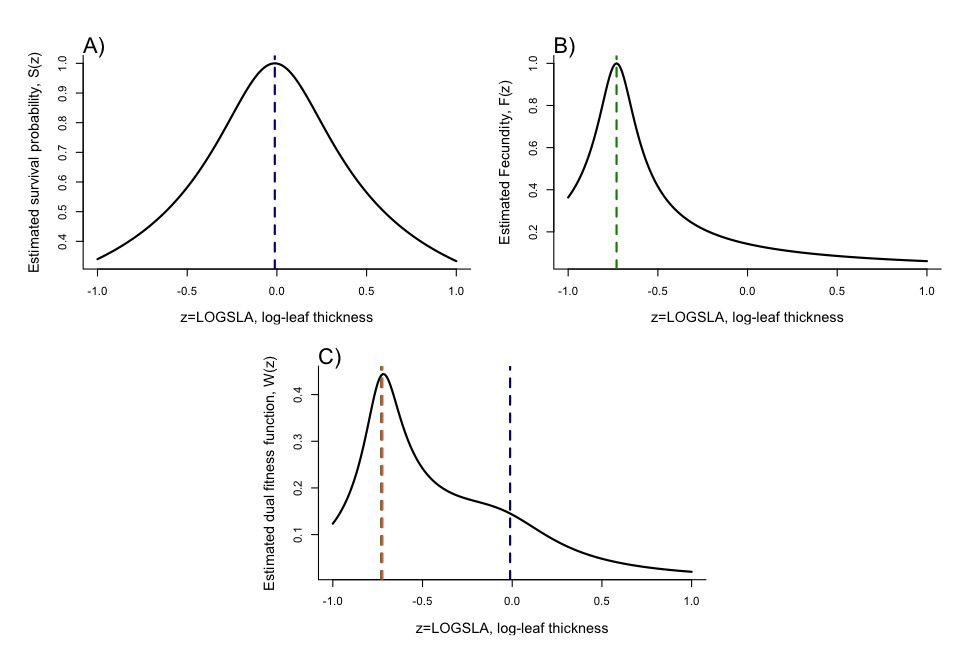}
\caption{A) Estimated survival function (second term in equation \ref{heterofit}) under heterogeneity of the strength of stabilizing selection on log leaf thickness using data from Etterson (2004) and Etterson and Shaw (2001) for \textit{Chamaecrista fasciculata}, provided under package aster (Geyer 2025). The parameter estimates for our survival probability as a function of the trait were estimated via maximum likelihood under a Bernoulli probability model: $\hat{a}_S=0.4746$, $\hat{b}_S=0.0560$, $z_S^*=-0.01167$. B) Estimated fecundity function (first term in equation \ref{heterofit}) under heterogeneity of the strength of stabilizing selection on log leaf thickness using the same data set. The parameter estimates for our fecundity probability as a function of the trait were estimated via maximum likelihood under a normal approximation to a Poisson probability model of the observations: $\hat{a}_F=0.4986$, $\hat{b}_F=0.00549$, $z_F^*=-0.7301$. The nuisance parameter $\beta$ was estimated to be $\hat{\beta}=148.5948$. C) The estimated overall fitness function under heterogeneity of the strength of stabilizing selection (equation \ref{heterofit}) on log-leaf thickness for the same data set. Besides marking the two fitness component maxima, the overall fitness is marked with a dashed orange line at $-0.7237$}.\label{chamaesurv}
\end{figure}

As a proof of concept, we illustrate our general fitness model with random effects by fitting our fitness function model to the data from \cite{etterson2001constraint} and \cite{etterson:2004}, who provide both survival and reproduction data for the plant \textit{Chamaecrista fasciculata}. Because of the symmetric nature of the model for reproduction and survival (\textit{i.e.} the equation for the fitness is the product of two identical in form kernels of non-central t-distributions), care must be exerted when fitting this model to data.  Explicit labeled data for reproduction and for survival must be available to avoid label switching-type identifiability issues \citep{Rannala:2025}. As a simple example, consider data $X_{1}=x_{1}, X_{2}=x_{2},\ldots,X_{n}=x_{n}$ sampled from a mixture of two normal distributions with parameters $\mu_1,\sigma^{2}_1$ and $\mu_2,\sigma^{2}_2$ with mixing proportions $\pi_1$ and $\pi_2 =1-\pi_{1}$. Then, the model parameters can be defined using the vector $\theta = (\pi_1,\mu_1,\sigma^{2}_1,\mu_2,\sigma^{2}_2)$ or the vector $\theta' = (\pi_2,\mu_2,\sigma^{2}_2,\mu_1,\sigma^{2}_1)$. Then, the likelihood function for both sets of parameters $\theta$ and $\theta'$ is thus identical, and hence these parameters are non-identifiable \citep{Rannala:2025}.  Thus, to fit our general fitness model we independently estimated: (i) the model parameters $(a_S,b_S,z^{\ast}_S)$ using the survival binary data and a Bernoulli likelihood function, and (ii) the model parameters $(a_F,b_F,z^{\ast}_F)$ using the fecundity data and a Log-Normal approximation to a Poisson probability model of the observations. The estimated fecundity and survival functions for all the data, regardless of block \citep[as shown in][]{etterson:2004}, are plotted in Figures \ref{chamaesurv} A) and \ref{chamaesurv} B), whereas the compound fitness function is plotted in Figure \ref{chamaesurv} C). Note the remarkable resemblance between the estimated fitness function in Figure \ref{chamaesurv} C) and the plots in Figure \ref{Fig2} done with arbitrarily chosen parameters. Of course, these plots are done here just as a proof of concept, and formal model selection tests to confront different hypotheses and examine model adequacy should be carried out \citep{Taper:2008}, something that we leave for future developments. In the Supplementary Material I, we provide code to compute parametric bootstrap confidence intervals on the estimates.

\subsection*{Effect of the adaptive topography on the evolution of the mean phenotype}
\subsubsection*{Deterministic evolution}
When Lande (1976) assumed a single-component fitness function centered at zero, $W(z) = {\rm e}^{-\frac{z^{2}}{2\omega^{2}}}$, the form of the fitness function suggested that the strength of selection and the trait optimum were unrelated to one another. Our above reconsideration of the same process but with multiple, non-zero optima for survival and fecundity fitness shows that the trait optimum and the strength of selection may be necessarily intertwined, a point that was obscured in Lande's original development and seems to remain widely underappreciated to this day. Although Lande's adaptive topography equation still holds with our modified fitness function under constant selection in space and time, the interpretation of the selection force is crucially changed. 

To study this change, first note that the average phenotype after selection from \cite{Lande:1976} is: 
\begin{equation}\label{aphas}
\Wb{z}_{W}(t) =  \frac{\omega^{2}}{\sigma^{2}+\omega^{2}} \Wb{z}(t)
\end{equation}
\noindent  (see equation \ref{fullaphas} Appendix \ref{ALandemath}). As $\omega^2/(\sigma^2 + \omega^2)$ is bounded between zero and one, equation \ref{aphas} states that the average trait value following selection is only a fraction of its value before selection. When we plug the fitness function written with both survival and reproduction components (equation \ref{newfitness}) into equation (\ref{intaphas}), and the integrands are simplified using Result 1 (Appendix \ref{AGausskernel}), the resulting expression for $\Wb{z}_{W}(t)$ becomes:
\begin{equation}\label{newaphas}
\Wb{z}_{W}(t) = \frac{\omega^{2}}{\sigma^{2}+\omega^{2}}\bigg[ \Wb{z}(t) + \sigma^{2}(\delta z_{F}^{\star} + \gamma z_{S}^{\star}) \bigg],
\end{equation}
\noindent (full derivation available in Appendix \ref{Azwbarmath}).

To further compare Lande's results with our results assuming the two-fitness components function (equation \ref{newfitness}), in what follows we will denote Lande's strength of stabilizing selection as $\omega^{2}_{0}$ and ours simply as $\omega^{2} = \frac{1}{\delta + \gamma}$.  While Lande's force of selection simplifies to
\begin{equation}\label{Linfmean}
\Wb{z}(t+1) - \Wb{z}(t) = -\frac{\sigma^{2}}{\omega^{2}_{0}+ \sigma^{2}} h^{2}\Wb{z}(t), 
\end{equation}
\noindent in the Appendix \ref{Azwbarmath}, we show that when our dual components fitness function is used, we obtain
\begin{align}\label{Ourinfmean}
\Wb{z}(t+1) - \Wb{z}(t) &= \frac{h^{2}\sigma^{2}}{\omega^{2} + \sigma^{2}}\left(\omega^{2}[\delta z_{F}^{\star} + \gamma z_{S}^{\star}] - \Wb{z}(t)\right) \nonumber\\
\nonumber \\
&=\frac{h^{2}\sigma^{2}}{\omega^{2} + \sigma^{2}}\left(\mathcal{P}z_{F}^{\star} + (1-\mathcal{P})z_{S}^{\star} -\Wb{z}(t)\right).
\end{align}
\noindent Thus, using this dual components fitness function renders obvious that the ``adaptive zone'' (\textit{i.e.}, the derivative of the average fitness with respect to the average phenotypic value) bears a dependency between the optimum trait value and the overall strength of stabilizing selection $\omega^{2}$. At equilibrium, the mean trait value is between the optima of each fitness component, and closer to the optimum of the trait with stronger selection. Importantly, this dependency persists through the formulation of the OU model of evolution, as we will show next.


\subsubsection*{Effect of random genetic drift on evolution}

Lande's derivation of the OU model proceeds from equation \ref{Linfmean} by moving from a deterministic, discrete-time difference equation for that force of selection to a stochastic discrete-time process, which is more amenable to statistical model fitting. This stochastic process considers the dynamics of the time-dependent random variable $\bar{Z}(t)$, representing the distribution of possible mean trait values at any given time, $t$. To describe the dynamics of the random variable, one requires the first and second moments of the distribution of the change in mean fitness, which are
\begin{align}
    {\rm E}[\Delta \Wb{Z}(t)|\Wb{Z}(t) = \Wb{z}(t)] &= -\frac{\sigma^{2}h^{2}\Wb{z}(t)}{\omega^{2}_{0}+ \sigma^{2}}
\end{align}
\noindent and
\begin{align}
    {\rm E}[(\Delta \Wb{Z}(t))^{2}|\Wb{Z}(t) = \Wb{z}(t)] &= {\rm Var}[\Delta \Wb{Z}(t)|\Wb{Z}(t) = \Wb{z}(t)] + \left\{{\rm E}[\Delta \Wb{Z}(t)|\Wb{Z}(t) = \Wb{z}(t)]\right\}^{2}\nonumber\\
    \nonumber\\
    &\approx \frac{h^{2}\sigma^{2}}{N},
\end{align}
\noindent respectively, where $N$ is a large random sample size of the average phenotype of offspring of selected individuals,  and it is assumed that $\omega^{2}_{0} >> h^{2}\sigma^{2}$, a statement that the expected change in the mean should be ``small'' in any unit of time. 

The above moments for the dynamics of trait change can be scaled to a diffusion process, for which many probabilistic and statistical fitting tools are available. A commonly used and flexible model of stochastic dynamics over time is the Ornstein-Uhlenbeck (OU) model, which is often written as
$$
dX(t) = \alpha (\theta - X(t))dt +  \sqrt{\tau^{2}}dB(t).
$$
In this form, $\theta$ is the long-term mean of the process $X(t)$, $\alpha$ is the rate of approach to the mean, $\tau^2$ is the variance of the process over some small unit of time, and $B(t)$ is a random variable following the properties of Brownian motion. Lande scaled the dynamics of $\bar{Z}$ to just such a diffusion process with infinitesimal mean and variance matching the moments above, and thus defined the OU process for the mean trait value as 
$$
d\bar{Z}(t) = -\alpha_0 \bar{Z}(t)dt +  \sqrt{\tau^2}dB(t),
$$
\noindent where $\alpha_0 = \frac{\sigma^{2}h^{2}}{\omega^{2}_{0}+ \sigma^{2}}$ came to be known as ``the strength of selection'' and $\tau^2 = \frac{h^{2}\sigma^{2}}{N}$ is the sampling variance of the average phenotype. 

\cite{Hansen:1997} and then \cite{Butler:2004} generalized this model by formulating a process where the trait optimum was given an arbitrary non-zero value, $\theta_{0}$, thus changing the infinitesimal mean of the OU process to $\alpha_{0}(\theta_{0}-x)$, which results in the equation
\begin{equation}\label{BK2004}
d\bar{Z}(t) = \alpha_{0}(\theta_{0} - \bar{Z}(t))dt +  \sqrt{\tau^{2}}dB(t).
\end{equation}
\noindent Perhaps one of the reasons why this model has been so successful and widely used in the literature is that it is easy to simulate data from it.  
\noindent Since Hansen's seminal paper, the field of comparative methods in phylogenetics has seldom stopped to look back at the mechanisms yielding that optimum, $\theta_{0}$, and rather focused on comparing single vs multiple optima models of evolution along a phylogeny. However, note that using our new fitness function definition, equation (\ref{newfitness}), and the resulting first moment of the change in the average phenotype, the OU model of trait evolution has mean $\theta = \omega^{2}(\delta z_{F}^{\star} + \gamma z_{S}^{\star})$ and strength of selection $\alpha = \frac{h^{2}\sigma^{2}}{\omega^{2}+\sigma^{2}}$. With this result, the dependency between the optimal trait, $\theta$, and the strength of selection, $\alpha$, is rendered explicit and calls for the study of the implications of this dependency. The corresponding stochastic differential equation for the average size of a trait then becomes:
\begin{equation}\label{newdiffusion}
\begin{array}{lll}
d\bar{Z}(t) &=& \alpha(\theta - \bar{Z}(t))dt +  \sqrt{\tau^{2}}dB(t)\\
 & & \\
 &=& \frac{h^{2}\sigma^{2}}{\omega^{2}+\sigma^{2}}(\omega^{2}(\delta z_{F}^{\star} +\gamma z_{S}^{\star}) - \bar{Z}(t))dt +  \sqrt{\tau^{2}}dB(t)\\
 & & \\
  &=& \frac{h^{2}\sigma^{2}}{\omega^{2}+\sigma^{2}}(\mathcal{P}z_{F}^{\star} + (1-\mathcal{P})z_{S}^{\star} - \bar{Z}(t))dt +  \sqrt{\tau^{2}}dB(t)
\end{array}
\end{equation}

Crucially, the amount of stochastic exploration of the adaptive landscape by the average phenotype $\bar{Z}(t)$ is now altered by this dependence of the infinitesimal mean on the strength of stabilizing selection.  The exploration around a region $\mathcal{R}$ for a stochastic process attaining a stationary distribution $\pi(\bar{z})$ is explicitly given by $\lim_{t\rightarrow \infty}P(\bar{Z}(t) \in \mathcal{R})=\int_{\mathcal{R}}\pi(\bar{z})d\bar{z}$. For the Butler and King process, at stationarity, $\bar{Z}_{\infty} \sim {\rm Normal}(\theta_{0}, \frac{\omega^{2}+\sigma^{2}}{2N})$ is normally distributed with mean independent of the variance, whereas for our version of the OU process, at stationarity, $\bar{Z}_{\infty} \sim {\rm Normal}(\omega^{2}(\delta z_{F}^{\star} + \gamma z_{S}^{\star}), \frac{\omega^{2}+\sigma^{2}}{2N})$ is normally distributed for which the mean and the variance have shared, mutual dependency on $\omega^2$. Simply ignoring such shared dependency while still attempting to fit the Butler and King OU model will not solve estimability problems for this model parameters. Rather, we suggest adding whenever possible information regarding the component's individual optima, all derived from field or laboratory experiments and observations. Note that from equation (\ref{thetaweights}) the optimal trait value is a weighted average of the optimal trait values for fecundity and survival components of fitness. One can then study the change in the OU optima as the relative strengths of stabilizing selection on different fitness components change, as well as how the individual fitness component optima might change.


Because the weights of the average of fitness components are given by the relative strength of selection on each component, the overall strength of selection on a trait is inextricably linked to the optimal values for the components. This heretofore unrecognized point provides a potential explanation whenever $\theta_{0}$ proves to be difficult to estimate separately from $\alpha_{0}$ in Butler and King's OU process. Indeed, ridged bivariate likelihoods along the $\theta_{0}$ and $\alpha_0$ axes seem to be common \citep{grabowski:2023}, a clear indication of underlying joint dependency of two model parameters. Under Butler and King's OU process, a change in a trait optimum occurs independently of any change in the strength of stabilizing selection, whereas in our case, such a change may or may not occur independently of changes in stabilizing selection. 

\subsubsection*{Stochastic dynamics under temporal and spatial heterogeneities}
Heterogeneity in the strength of stabilizing selection in reproduction, survival, or both is but only one of the potential hypotheses that could be advanced to study the evolution of asymmetry in fitness functions. Yet, by itself, this hypothesis leads to a vast family of fitness functions, which in turn lead to a family of diffusion models of evolution. In this section, we develop two general diffusion models representing two distinct classes of models within that vast family of possible diffusion processes, which result from two different modalities of implementing heterogeneity in the strengths of stabilizing selection.


In what follows, we construct sample models via simulations that incorporate both types of heterogeneity. Annotated R code reproducing simulations and Figures \ref{Fig6} and \ref{Fig7} is provided in the Supplementary Material II and III. To facilitate comparison among alternative fitness landscapes, the diffusion variance was held constant across all simulations,
\(
\tau^2=\frac{h^2\sigma^2}{N}.
\)
Consequently, stochastic fluctuations in the population mean phenotype are identical among models, and differences in evolutionary trajectories arise solely from differences in the deterministic component of selection. In both cases, the basic steps to derive the diffusion process are the same, namely:

\begin{enumerate}
    \item[\textbf{Step 1}] Define the new fitness function $W(z)$, starting from equation (\ref{newfitness}). If considering spatial heterogeneity in the strength of stabilizing selection, start with the fitness function in equation (\ref{heterofit}). To do that, specify the probability distribution(s) for $\delta$ and/or $\gamma$ and integrate $W(z)$ over this(these) source(s) of variability. 
    \item[\textbf{Step 2}] With the new fitness function in place, compute the average trait value after selection 
    $$\Wb{z}_{W}(t) = \frac{\int z W(z) p(z,t)dz}{\int W(z) p(z,t)dz} =  \frac{A}{B}.$$
    To calculate these integrals, assume a distribution of the trait at time $t$. Lande used a normal density with mean $\bar{z}(t)$ and variance $\sigma^{2}$. Note that Lande's connection between the breeder's equation and the rate of change of the average fitness (equation \ref{LAT}), which is analogous to Wrights' equation for gene frequencies, depends only on the normality assumption of $p(z,t)$ but not on the particular fitness function form used  \citep[see][eqs. 6 and 7]{Lande:1976}. Note that steps 1 and 2 are the same if one wishes to make only a deterministic change to the mean phenotype. 
    \item[\textbf{Step 3}] The new diffusion process of a trait will have an infinitesimal mean and variance: 
    \[
    m(z) = (\Wb{z}_{W}(t)- \bar{z}(t))h^{2}, \qquad \tau^{2} = \frac{h^{2}\sigma^{2}}{N},
    \]respectively. The drift term, $m(z)$, is obtained in Step 2 from the breeder's equation, whereas the diffusion term, $\tau^2$, follows from Lande's diffusion approximation.
\end{enumerate}

For spatial heterogeneity, these steps yield a single effective fitness landscape obtained by averaging over heterogeneous environments. Operationally, this is accomplished by specifying distributions for $\delta$ and/or $\gamma$ and integrating the fitness function over these sources of variability before computing the selection gradient. The resulting fitness landscape can therefore be interpreted as the average selective environment experienced by the population. In general, this effective fitness function is non-Gaussian and the corresponding selection gradient does not admit a closed-form expression. Consequently, the selection gradient and the corresponding drift function $m(z)$ are evaluated numerically and then used to simulate the diffusion process in Step 3.

By contrast, under temporal heterogeneity, variation in the strengths of stabilizing selection occurs through evolutionary time across generations rather than across environments. In this case, the fitness function is initially computed as the product of the Gaussian fecundity and survival kernels in equation (\ref{newfitness}), and the breeder's equation is used to derive the corresponding diffusion process. Randomness in $\delta$ and $\gamma$ is then introduced through time by repeatedly sampling new values of these parameters during the simulation. Consequently, both the optimum phenotype and the strength of attraction become time-dependent random variables. Specifically, the instantaneous optimum is
\[
\theta(t)
=
\frac{\delta(t)z_F^\star+\gamma(t)z_S^\star}
{\delta(t)+\gamma(t)},
\]
whereas the strength of attraction is determined by the total strength of stabilizing selection,
\[
\delta(t)+\gamma(t).
\]

The population mean phenotype therefore evolves on a sequence of temporally fluctuating fitness landscapes. In our simulations, the mean phenotype was evolved using the Euler--Maruyama method to solve stochastic differential equations \citep{Dennis:2014} while repeatedly updating the selection parameters through time. Compared to a constant-parameter Ornstein-Uhlenbeck process, temporal heterogeneity introduces temporal variation in both the optimum phenotype, $\theta(t)$, and the strength of attraction toward that optimum. Because the optimum changes through time, the mean phenotype may transiently lag behind the instantaneous optimum. In contrast, spatial heterogeneity produces a single effective fitness landscape obtained by averaging over heterogeneous environments, so the resulting evolutionary dynamics occur on a fixed landscape rather than a moving optimum.


\begin{figure}[htbp]
\centering
\includegraphics[angle=0,width=1\textwidth]{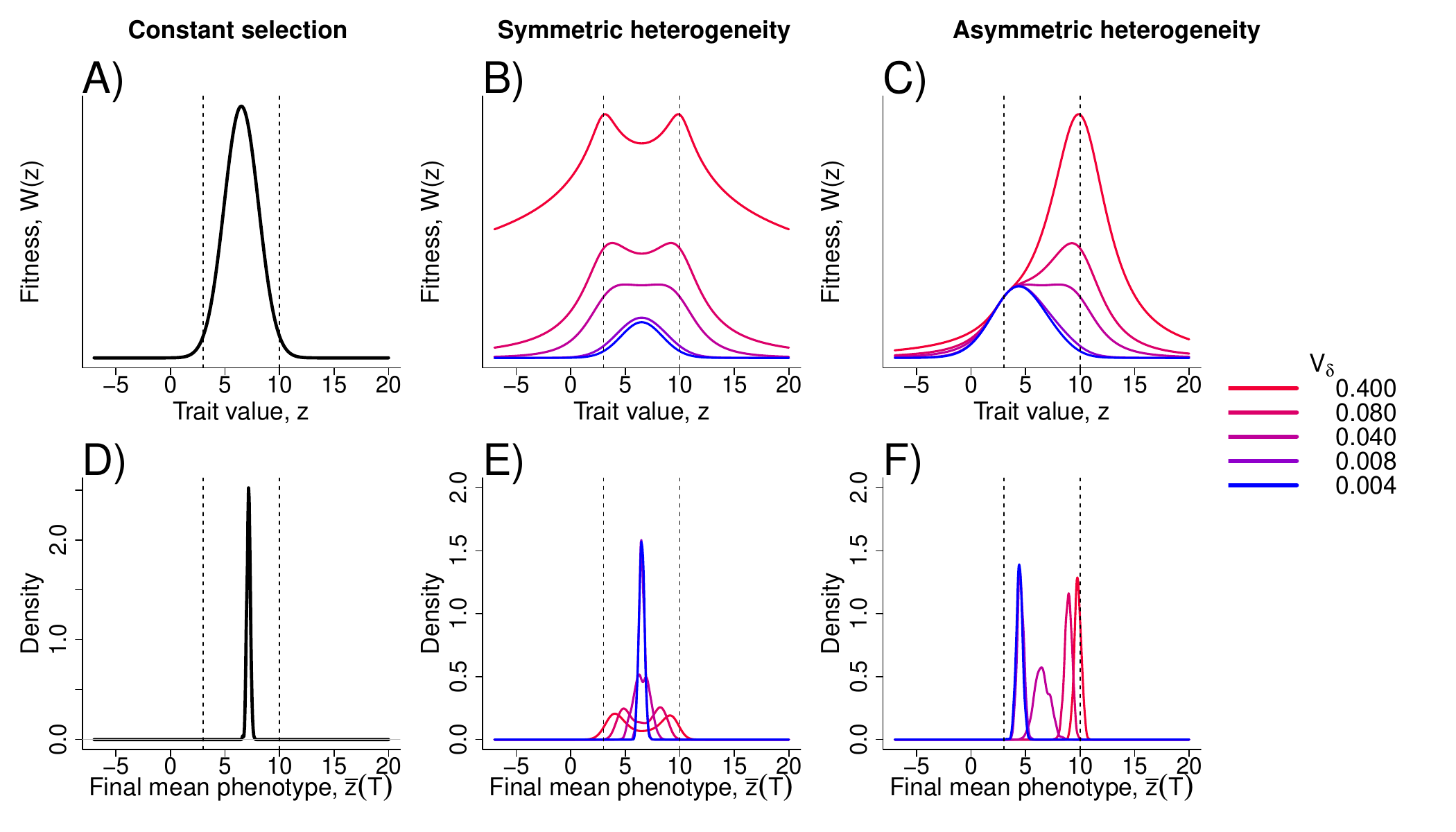}
\caption{Effects of spatial heterogeneity in selection strength on fitness landscapes and long-term evolutionary outcomes. Top row: effective fitness landscapes under constant selection (A), symmetric spatial heterogeneity (B), and asymmetric spatial heterogeneity (C). Dashed vertical lines indicate the phenotype maximizing fecundity ($z_F^{\star}=3$) and the phenotype maximizing survival ($z_S^{\star}=10$). Under constant selection, fecundity and survival selection strengths are fixed at $\delta=\gamma=0.2$. Under spatial heterogeneity, the mean selection strengths are held constant at $E[\delta]=E[\gamma]=0.2$, while the variances are varied according to $V[\delta]=V[\gamma]\in\{0.400,0.080,0.040,0.008,0.004\}$. Colors range from red (high heterogeneity) to blue (low heterogeneity). In the symmetric case, heterogeneity is varied simultaneously in both fecundity and survival selection. In the asymmetric case, heterogeneity is varied only in fecundity selection ($\delta$), while survival selection remains fixed at $\gamma=0.2$. Bottom row: kernel density estimates of the distribution of the final mean phenotype, $\bar z(T)$, obtained from 1000 independent simulations for each parameter combination. Simulations used $N=50$, $V_P=1.5$, $h^2=(V_P-1)/V_P$, $\tau=h^2V_P/N$, $\bar z(0)=6.5$, and $T=200$, $a_F={0.1,0.5,1,5,10}$, $b_F={0.5,2.5,5,25,50}$, $a_S=a_F$ in B or $a_S=1$ in C, $b_S=b_F$ in B or $b_S=5$ in C.}\label{Fig6}
\end{figure}

Recall that when we first defined temporal heterogeneity, we gave an example in which we let $\delta$ and $\gamma$ each be Gamma distributed with the same rate parameter $b$. Then, the relative contribution of fecundity selection, $\mathcal{P}$, may be assumed to be Beta distributed with shape parameters equal to $\alpha_F$ and $\alpha_S$, whereas the total strength of stabilizing selection,
\(
T(t) = \delta(t)+\gamma(t),
\)
is Gamma distributed. The instantaneous optimum can therefore be written as $\theta(t)=\mathcal{P}(t)z_{F}^{\star} + (1-\mathcal{P}(t))z_{S}^{\star}$. Under this representation, temporal variation in $\mathcal P(t)$ determines the location of the optimum phenotype, while temporal variation in $T(t)$ determines the strength of attraction toward that optimum. Thus, the Beta-Gamma decomposition separates fluctuations in the position of the optimum from fluctuations in the overall strength of stabilizing selection. This formulation provides a convenient framework for exploring alternative hypotheses regarding the relative importance of each fitness component and their variability.   

A more general approach to incorporating temporal heterogeneity would be to specify $\theta$ via another parallel stochastic differential equation (SDE) process evolving according to a second SDE. For example, with the $\theta$ parameter varying randomly over time according to a diffusion process with a beta transition probability distribution and a beta stationary distribution, one may posit different hypotheses regarding the form of the relative variation and importance of the strengths of stabilizing selection in reproduction and survival.


Figure \ref{Fig6} illustrates how the effective fitness landscape changes as the variance in selection strength is varied while maintaining a constant mean selection strength and spatial heterogeneity. For the particular combinations of $a_i$ and $b_i$ with $i\in\{F,S\}$ considered here, increasing heterogeneity in both reproduction and survival produces increasingly broad fitness landscapes with multiple local peaks, while preserving overall symmetry around the midpoint between $z_F^\star$ and $z_S^\star$ (Figure \ref{Fig6}B). In contrast, when heterogeneity is varied only in reproduction across generations, the fitness landscape becomes increasingly asymmetric and its dominant peak shifts toward the survival optimum (Figure \ref{Fig6}C). Thus, for these parameter values, heterogeneity modifies not only the width of the fitness landscape but also its overall shape and the location of its principal fitness peak.

The distributions of endpoint phenotypes reflect these changes in landscape structure (Figure \ref{Fig6}, panels D-F). Under symmetric heterogeneity, increasing variance broadens the distribution of final mean phenotypes and reduces concentration around a single evolutionary outcome, consistent with the increasingly diffuse landscapes shown in Figure \ref{Fig6} panels B and E. Under asymmetric heterogeneity, increasing variance shifts the endpoint distributions toward larger trait values, mirroring the displacement of the dominant fitness peak (Figure \ref{Fig6}, panels C and F).

\begin{figure}[htbp]
\centering
\includegraphics[angle=0,width=1\textwidth]{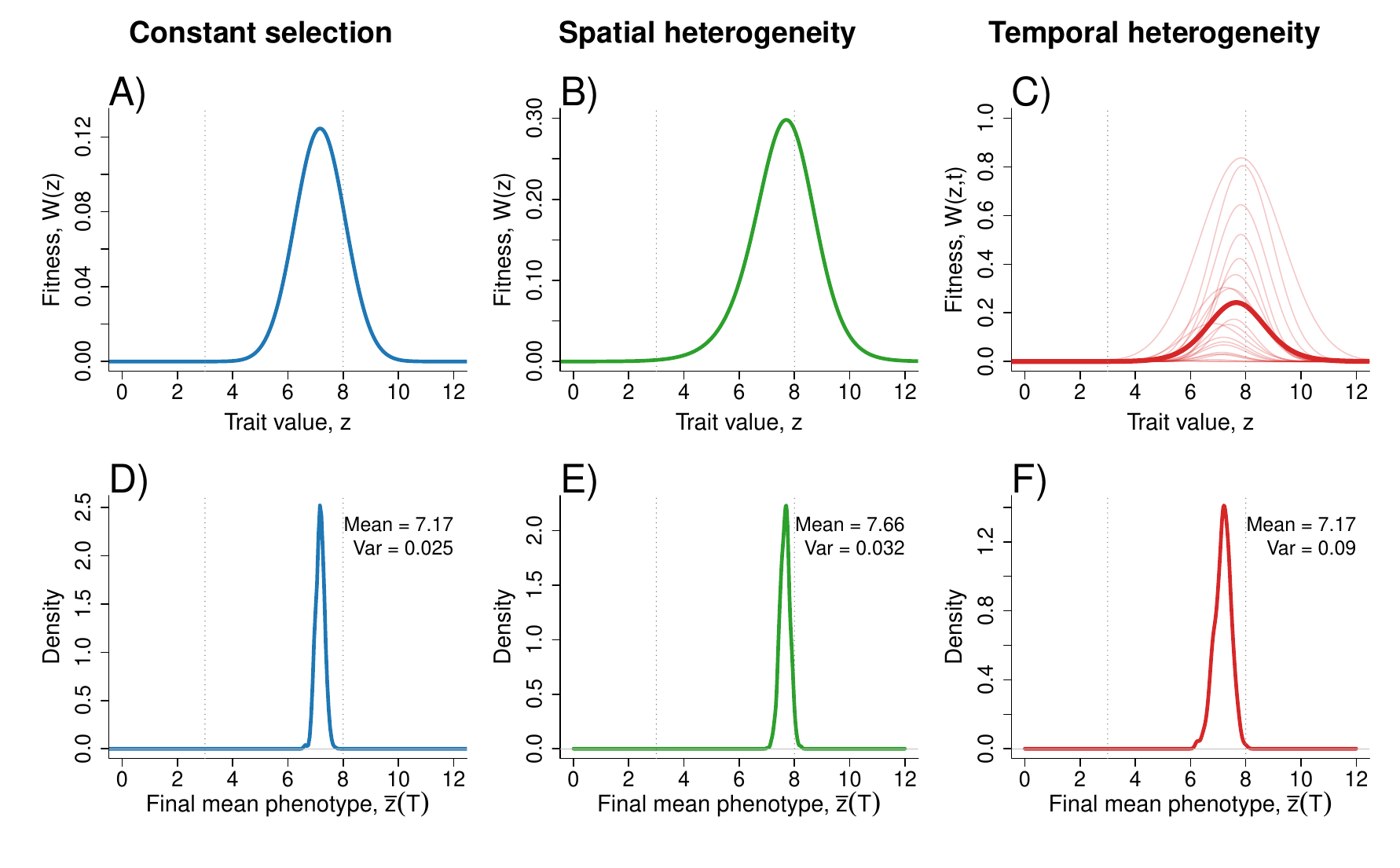}
\caption{Fitness landscapes and long-term evolutionary outcomes under constant selection, spatial heterogeneity, and temporal heterogeneity. Top row: fitness landscapes under constant selection (A), spatial heterogeneity (B), and temporal heterogeneity (C). Dashed vertical lines indicate the phenotype maximizing fecundity ($z_F^{\star}=3$) and the phenotype maximizing survival ($z_S^{\star}=8$). Under constant selection, fecundity and survival selection strengths are fixed at $\delta=0.2$ and $\gamma=1$, respectively. Under spatial heterogeneity, selection strengths are Gamma distributed with $\delta\sim\Gamma(1,5)$ and $\gamma\sim\Gamma(5,5)$, producing a non-Gaussian spatially averaged fitness landscape. Under temporal heterogeneity, thin red curves represent instantaneous fitness landscapes generated by independently sampling $\delta$ and $\gamma$ through time, whereas the thick red curve shows their empirical time average. The average landscape is displayed for visualization only and is not used directly in the evolutionary simulations. Bottom row: kernel density estimates of the distribution of the final mean phenotype, $\bar z(T)$, obtained from 1000 independent simulations. Simulations used $N=50$, $V_P=1.5$, $h^2=(V_P-1)/V_P$, $\tau=h^2V_P/N$, $\bar z(0)=5.5$, and $T=200$.}\label{Fig7}
\end{figure}

Figure \ref{Fig7} compares the effective fitness landscapes and long-term evolutionary outcomes generated by constant selection, spatial heterogeneity, and temporal heterogeneity. To facilitate comparison, the same diffusion coefficient was used in all three models, so differences among outcomes arise from differences in the deterministic component of selection. For the heterogeneous models, the distributions of selection strengths were chosen such that $\delta \sim \Gamma(1,5)$ and $\gamma \sim \Gamma(5,5)$. Consequently, the spatial and temporal models share the same underlying distributions of selection strengths but differ in how heterogeneity is incorporated: spatial heterogeneity averages over variability to produce a single effective fitness landscape, whereas temporal heterogeneity generates a sequence of temporally fluctuating fitness landscapes.

The endpoint distributions are broadly consistent with the corresponding fitness landscapes. Constant selection and temporal heterogeneity produce similar mean endpoint phenotypes, indicating that temporal fluctuations in selection strength have relatively little effect on the average long-term evolutionary outcome. However, temporal heterogeneity generates a substantially broader distribution of endpoint phenotypes, reflecting the greater variability introduced by a continually changing optimum. In contrast, spatial heterogeneity shifts the effective fitness landscape toward the survival optimum and produces a corresponding shift in the distribution of endpoint phenotypes. Thus, spatial heterogeneity primarily alters the location of long-term evolutionary outcomes, whereas temporal heterogeneity primarily increases their variability.


Just like the Little Prince's rose that always wanted more (Chapitre VIII, ``Le petit Prince''), biologists have a strong penchant for making any and most models as complex as possible.  The natural ``want'' in this case is to incorporate both spatial and temporal heterogeneities. Doing so can of course be done, the question is \textit{should} adding complexity be done and why? Start, for instance, with the fitness function in equation (\ref{newfitness}) and proceed with the steps above. Then, to add temporal heterogeneity on top of spatial heterogeneity, ``simply'' elicit an SDE model for one or more of the parameters $a_{i},b_{i}$. The price for so doing, however, may be quite steep: if estimation of the simple OU process is known to be difficult and data hungry, adding a few extra parameters and complexities will not necessarily improve the estimation dilemmas. Alas, probing at the adequacy of one particular model remains a non-trivial task in and of itself that we believe should be done according to the model selection strategy of \cite{Taper:2021} and \cite{Edgar:2022} to quantify local vs global model uncertainties.

\section*{Discussion}
Mathematical models have long guided our understanding of the process of phenotypic evolution. It is not an exaggeration to state that these models carry the meaning of our scientific guild \citep{Taper:2008}. Hence, it is of paramount importance to constantly question fundamental aspects upon which our theories are founded. Despite being fifty years old, \cite{Lande:1976}'s creative derivation of the adaptive phenotypic topography still conceals implicit assumptions that warrant scrutiny. In particular, Lande did not decouple reproduction and survival in his formulation of the fitness function, yet the same organismal trait could affect these fitness components in different ways. Thus, all the ensuing formulae and results of his work, from the expression of the average phenotype after selection and the average phenotypic response to selection in one generation to the diffusion approximation for the evolution of the mean trait value, carry the weight of ignoring this decoupling. Here, we show that both the location of the integrative optimum (an adaptive compromise between two fitness components) and the overall maximum of the fitness function are modulated by the strengths of stabilizing selection for reproduction and survival. Specifically, the overall optimum is cast \textit{de novo} as a weighted average of the optima in reproduction and survival. Interestingly, the weights are given by the relative strengths of stabilizing selection on each fitness component. In addition, absolute fitness is reduced by discordant reproductive and survival optima. The magnitude of this reduction depends jointly on the distance separating the two optima and on the strengths of stabilizing selection acting on reproduction and survival. Thus, the decoupling of fitness into reproductive and survival components affects not only the location of the adaptive optimum but also the overall elevation of the fitness landscape. While the reduction in absolute fitness plays no role directly in the expected trajectory of the mean phenotype along the fitness landscape, it can indirectly influence selection through changes in population size and the overall efficacy of selection in relation to non-selective factors. Future efforts to estimate the integrative optimum of a phenotypic trait should attempt to gather independent information regarding the strengths of selection in reproduction and survival.

The explicit decoupling of survival and reproduction in the fitness function ultimately leads to a previously unrecognized strategy or framework to derive an entire family of stochastic differential equation models of the evolution of the mean phenotype. These models provide alternative models to the famous OU model of evolution. Besides extensions and modifications to the OU model of phenotypic evolution \citep{Hansen:1997,Butler:2004, Uyeda:2011, Pennell:2014, Uyeda:2014, Uyeda:2015, Uyeda:2018,  Blomberg:2020, Week:2021, Boyko:2023}, recent efforts notably include generalizations that can accommodate any shape in macroevolutionary landscapes \citep{Boucher:2018} but the field is marked by a paucity of alternative models embodying different hypotheses regarding the evolution of continuous traits along a phylogeny. Our strategy mirrors closely ideas from stochastic population dynamics models. In particular, spatial and temporal variabilities in selection mirror stochastic models of demographic and environmental stochasticities \citep{Ponciano:2018} but differ in key particular ways, like the scale and shape of the infinitesimal variance.

Interlocking the location of the optimum with the strengths of stabilizing selection substantially changes the behavior of the subsequent stochastic OU process of the average phenotype. Importantly, well-known quantities that characterize diffusion processes, like the amount of stochastic exploration of the adaptive landscape by the average phenotype $\Wb{Z}(t)$ in a region of the state-space $\mathcal{R}$, or mean first passage times (the average time it takes the process to go from a value $z_0$ to a level $y$), should not any more be assumed to be modulated independently by either changes in the optimum location conceived as a stand alone parameter, or by changes in the strengths of stabilizing selection. Unlike any other Gaussian kernel, changing the strengths of stabilizing selection for reproduction and survival will induce a change in the location of the optimum.

A central consequence of heterogeneity in selection width is that the effective adaptive landscape becomes an emergent property of environmental variation itself. This contrasts with the homogeneous case, where the composite fitness landscape remains Gaussian and can be completely characterized by the component-specific optima and selection widths. Under heterogeneous conditions, however, both the shape of the landscape and its evolutionary consequences depend on how variation in selection width is distributed across space and time. Spatial heterogeneity primarily alters the geometry of the effective fitness landscape through averaging across environments, whereas temporal heterogeneity alters its dynamics by changing the relative contribution of fitness components through time. 

Spatial heterogeneity in selection strength generates emergent adaptive landscapes whose properties cannot be inferred directly from any single local fitness function. Even when all component fitness functions are Gaussian, averaging across heterogeneous environments produces effective landscapes that are non-Gaussian and often non-quadratic. Consequently, population-level descriptions of selection may differ substantially from local descriptions of selection. Viewed in this light, failures of quadratic approximations may sometimes reflect properties of the underlying fitness landscape rather than limitations of the statistical models used to describe it \citep{mitchell:1987, shaw:2010}. Previous work demonstrated that multiplicative fitness components can generate complex fitness surfaces and complicate statistical inference \citep{mitchell:1987, shaw:2008, shaw:2010}. Our results extend this perspective by showing that environmental heterogeneity in selection strength can generate non-Gaussian composite landscapes even when all local fitness functions remain Gaussian. In addition to altering the location and breadth of adaptive peaks, heterogeneity can substantially modify the overall height of the effective fitness landscape. In particular, unequal amounts of heterogeneity in reproduction and survival generate asymmetric landscapes whose maxima may be considerably elevated or depressed relative to the symmetric case. Thus, spatial heterogeneity in selection strength influences not only where selection acts most strongly, but also the relative fitness advantage associated with different regions of phenotype space.

These changes in landscape geometry also have important consequences for the maintenance of phenotypic diversity. Several mechanisms have been proposed to explain the persistence of phenotypic variation despite stabilizing selection, including environmental heterogeneity, trade-offs among fitness components, temporally fluctuating selection, and eco-evolutionary feedbacks linking ecological and evolutionary processes \citep{Kingsolver:2001, shaw:2010, govaert:2019, yamamichi:2023}. Our results highlight a specific role for spatial heterogeneity in selection strength. By broadening effective fitness landscapes and weakening the fitness penalties associated with moderate maladaptation, heterogeneous selection strengths can increase the range of phenotypes capable of maintaining relatively high fitness. Such effects provide a potential mechanism through which environmental heterogeneity contributes to the persistence of phenotypic variation despite stabilizing selection.

Importantly, spatial heterogeneity in selection strength may influence not only the amount of phenotypic variation that is maintained, but also its distribution. Because averaging across environments with different selection strengths can generate asymmetric effective fitness landscapes, signatures of spatial heterogeneity may be detectable in the shape of phenotypic distributions themselves. Under mutation-selection balance, such asymmetric landscapes generate uneven distributions of standing genetic variation around the fitness optimum \citep{burger:1994}. When fitness declines more steeply on one side of the optimum than on the other, selection removes variation more efficiently from the steep side, even when mutational effects are approximately symmetric \citep{burger:1994,urban:2013}. Consequently, standing genetic variation tends to accumulate preferentially on the shallow side of the fitness landscape \citep{burger:1994}. When changes in environmental conditions shift the effective optimum, adaptive responses may therefore proceed more readily in directions associated with the shallow side of the landscape because greater standing genetic variation is expected to be maintained there \citep{burger:1994,vercken:2012}.

The consequences of heterogeneous stabilizing selection may extend beyond the maintenance of phenotypic variation to population persistence itself. Existing theory has emphasized the roles of local adaptation, spatial buffering, and adaptation along environmental gradients in heterogeneous environments \citep{kirkpatrick:1997,kawecki:2004,holt:2011,chevin:2011}. Our results suggest that the geometry of the effective fitness landscape may represent an additional and largely overlooked determinant of persistence. Understanding how these geometric effects interact with dispersal and demographic processes remains an important direction for future work.

The consequences of temporal heterogeneity in selection strength differ fundamentally from those of spatial heterogeneity. Rather than generating a static mixture of fitness surfaces, temporal variation in selection strengths continually alters the relative contribution of component fitness functions. As a consequence, both the effective optimum and the curvature of the adaptive landscape fluctuate through time even when component-specific optima remain fixed. Temporal heterogeneity therefore reshapes the dynamics of the landscape itself rather than simply altering its average geometry.

These dynamics connect naturally to the literature on fitness seascapes, which examines evolutionary change in temporally varying adaptive landscapes \citep{mustonen:2009}. In our framework, temporal variation in stabilizing-selection strength continually alters the relative contribution of fitness components to total fitness, causing both the location of the effective optimum and the curvature of the resulting landscape to fluctuate through time. Consequently, dynamic adaptive landscapes need not arise solely from changes in the selective targets associated with particular fitness components. Instead, temporal variation in selection strength can generate fitness-seascape dynamics by changing how fitness components are weighted. Effective optima therefore may move even when the underlying component-specific optima remain fixed, providing a process-based mechanism through which environmental variation can reshape adaptive landscapes through time.

Just as spatial heterogeneity modifies equilibrium phenotypic distributions by changing landscape geometry, temporal heterogeneity alters how populations track changing selective conditions. Fluctuations in selection strength generate intergenerational variation in effective optima and therefore in maladaptation. Consequently, populations may experience episodes of elevated mismatch even when the selective targets associated with reproduction and survival remain unchanged. Variation in selection width may therefore represent an underappreciated source of evolutionary variability capable of influencing adaptive tracking and responses to environmental change. Thus, temporal variation in selection width provides an alternative route through which fluctuating environments can influence adaptive tracking, complementing previous models that focused primarily on temporal variation in optima \citep{chevin:2013, chevin:2017,peniston:2020,peniston:2021,chevin:2025}.

Our results complement recent theoretical work emphasizing the role of multiple fitness components in determining adaptive compromises. In particular, \citet{cotto:2019} developed an elasticity-based framework in which multiple fitness components contribute to an integrative optimum through demographic weighting. In their framework, the relative importance of each component depends on demographic elasticities and therefore changes dynamically through eco-evolutionary feedbacks linking adaptation and population structure. By contrast, our model isolates the consequences of geometric weighting, in which the relative contribution of each component depends on the strength of stabilizing selection itself. Together, these studies suggest that adaptive compromises may be shaped by at least two distinct but potentially interacting processes: demographic weighting through elasticities and geometric weighting through selection strength. Integrating these mechanisms within a common framework may prove particularly important in temporally changing environments, where demographic feedbacks and fluctuations in selection strength are both expected to influence adaptive trajectories.

These considerations have direct implications for evolutionary rescue. Previous work has shown that temporal variation in the optimum of a single fitness function can either facilitate or hinder rescue depending on the temporal structure of environmental variation, particularly its autocorrelation \citep{chevin:2013, chevin:2017,peniston:2020,peniston:2021,chevin:2025}. In those models, environmental change acts primarily by shifting the optimal phenotype favored by selection. Our framework identifies a distinct mechanism. Temporal variation in selection strength alters the relative contribution of fitness components to total fitness and therefore generates fluctuations in the effective optimum even when the component-specific optima remain fixed. Consequently, populations may experience variation in maladaptation arising solely from changes in the weighting of competing selective pressures.

This effect is related to, but distinct from, recent results showing that random temporal variation in the width of a single fitness function can impede evolutionary rescue by increasing fluctuations in maladaptation (Godineau et al., in revision). In our model, temporal heterogeneity acts simultaneously on multiple fitness components and modifies the location of the effective optimum itself. The resulting rescue dynamics are therefore expected to depend not only on the magnitude and autocorrelation of width fluctuations, but also on how those fluctuations alter the weighting among competing component optima.

Finally, the evolutionary diffusion processes derived here may also have implications for macroevolutionary inference. Because the effective optimum and the strength of attraction toward that optimum are coupled, the resulting dynamics differ fundamentally from standard OU formulations. This coupling has direct implications for phylogenetic comparative methods. Standard OU models typically interpret changes in adaptive optima and changes in stabilizing-selection strength as distinct processes \citep{Hansen:1997, Butler:2004,Uyeda:2011}. In our framework, however, variation in selection strength can itself induce shifts in effective optima. Temporal variation in selection strength generates time-dependent adaptive landscapes, while spatial variation generates non-Gaussian effective landscapes. Both processes may therefore affect estimates of adaptive optima, evolutionary rates, and phylogenetic covariance structure. Determining whether phylogenetic data contain sufficient information to distinguish shifts in effective optima driven by changing selection strengths from shifts driven by changing selective targets represents an important challenge for future comparative analyses.

\section*{Conclusions}
Revealing and questioning implicit assumptions of foundational theories of evolutionary dynamics is para-mount to guiding the growth of our scientific field. By foregoing an explicit decoupling of the survival and reproduction fitness components, \cite{Lande:1976}'s landmark derivation of an adaptive topography concealed the fact that the integrative optimum is in fact a weighted average of both components' optima, and that the weights are the relative strengths of stabilizing selection for these components. While this result might be perceived at first blush as trivial, interlocking the location of the optimum with the strengths of stabilizing selection substantially changes our theoretical understanding of the malleability of the fitness function and the subsequent mathematical properties of the ensuing diffusion process for the average phenotype. In particular, our work revealed that expressing fitness as the product of two Gaussian kernels, one for reproduction and one for survival, and then allowing for integrating heterogeneity in the strengths of stabilizing selection endows the resulting fitness function with previously unrecognized flexibility. Our findings, therefore, suggest that empirical evidence for non‑Gaussian composite fitness landscapes might arise from variation in the width of selection functions across fitness components. This realization gave us a framework for deriving an entire family of stochastic differential equation models of the evolution of the mean phenotype. Because such flexibility and this modeling framework are tied to process variation in the strength of selection, they unlock unrealized potential to model and estimate, with empirical data, different ways in which the mean phenotype may evolve. Finally, understanding how selection widths vary across environments and through time may provide a unifying framework connecting adaptive-landscape theory, fitness seascapes, life-history evolution, evolutionary rescue, and macroevolutionary diversification.

\newpage

\renewcommand{\thesection}{\Alph{section}.\arabic{section}}
\renewcommand{\theequation}{\Alph{section}.\arabic{equation}}
\setcounter{section}{0}
\setcounter{equation}{0}
\section{Appendix}
\subsection{The product of two Gaussian kernels}\label{AGausskernel}

\begin{defn} The product of two Gaussian kernels with parameters $\mu_{i},\sigma^{2}_{i},\,i=1,2,$ is a weighted Gaussian kernel where the weight is given as a function of $\frac{(\mu_{1}-\mu_{2})^{2}}{2(\sigma^{2}_{1} + \sigma^{2}_{2}}$.  Specifically, 

\begin{equation}\label{Rslt1}
f(s) = {\rm e}^{-\frac{(s-\mu_{1})^{2}}{2\sigma^{2}_{1}}}{\rm e}^{-\frac{(s-\mu_{2})^{2}}{2\sigma^{2}_{2}}} = {\rm e}^{-\frac{(s-m^{\star})^{2}}{2\nu^{2}}},
\end{equation}
\noindent where 
$$
m^{\star} = \frac{(\sigma^{2}_{2}\mu_{1} + \sigma^{2}_{1}\mu_{2})}{\sigma^{2}_{1} + \sigma^{2}_{2}},\quad\text{and}\quad \nu^{2} = \frac{\sigma^{2}_{1}\sigma^{2}_{2}}{\sigma^{2}_{1} + \sigma^{2}_{2}}.
$$
\end{defn}
This result is arrived at by putting the exponent in $\exp\left\{-\left(\frac{(s-\mu_{1})^{2}}{2\sigma^{2}_{1}} + \frac{(s-\mu_{2})^{2}}{2\sigma^{2}_{2}}\right)\right\}$ under the same denominator, expanding, completing the square and simplifying.

\subsection{Lande's adaptive topography}\label{ALandemath}

Using Lande's single component fitness function $W(z) = \exp\left\{-\frac{z^{2}}{2\omega^{2}}\right\}$, the normal density $p(z,t)$ with mean $\Wb{z}(t)$ and phenotypic variance $\sigma^{2}$ and applying Result 1 above (equation \ref{Rslt1}) repeatedly, the average fitness reduces to
$$
\Wb{W} = \int W(z)p(z,t)dz  = \sqrt{\frac{\omega^{2}}{\omega^{2} +\sigma^{2}}}\exp\left\{- \frac{\Wb{z}(t)^{2}}{2(\omega^{2} +\sigma^{2})} \right\}.
$$
\noindent The expected value in the numerator of $\Wb{z_{W}}(t)$ integrates to
$$
\int z W(z)p(z,t)dz = \sqrt{\frac{\omega^{2}}{\omega^{2} +\sigma^{2}}}\left(\frac{\Wb{z}(t)\omega^{2}}{\sigma^{2}+\omega^{2}}\right){\rm e}^{-\frac{\Wb{z}(t)^{2}}{2(\omega^{2} +\sigma^{2})}}.
$$
\noindent Hence, the average phenotype after selection is simply 
\begin{equation}\label{fullaphas}
\Wb{z}_{W}(t) = \frac{{\rm E}_{Z}[ZW(Z)]}{{\rm E}_{Z}[W(Z)]} = \frac{\Wb{z}(t)\omega^{2}}{\sigma^{2}+\omega^{2}},
\end{equation}
\noindent Note also that $\frac{\partial \Wb{W}}{\partial \Wb{z}(t)} = \Wb{W}\left(-\frac{{\Wb{z}(t)}}{\omega^{2} + \sigma^{2}}\right)$. Then, equation \ref{changefit} is obtained upon multiplying and dividing by $\sigma^{2}$ and adding and subtracting $\Wb{z}(t)$ inside the parentheses; this derivative reduces to an expression containing the selection force:
\begin{align}
    \frac{\partial \Wb{W}}{\partial \Wb{z}(t)} &= \frac{\sigma^{2}}{ \sigma^{2}}\Wb{W} \left(-\frac{\Wb{z}(t)}{\omega^{2} + \sigma^{2}} + \Wb{z}(t) - \Wb{z}(t) \right)\nonumber\\
    &= \frac{\Wb{W}}{\sigma^{2}} \left(\frac{\omega^{2}\Wb{z}(t)}{\omega^{2} + \sigma^{2}} - \Wb{z}(t)\right)\nonumber\\
    &= \frac{\Wb{W}}{\sigma^{2}} \left(\Wb{z_{W}}(t) - \Wb{z}(t)\right).\nonumber
\end{align}
\noindent Solving for the selection differential above and substituting it into the breeder's equation, one indeed retrieves the adaptive topography equation (\ref{LAT}).

\subsection{The average phenotype after selection under a dual component fitness function}\label{Azwbarmath}

To compute the average phenotype after selection (equation \ref{intaphas})
\begin{equation}
\Wb{z_{W}}(t) = \frac{\int z W(z)p(z,t) dz}{\int W(z)p(z,t) dz},\nonumber
\end{equation}
under the dual components fitness function (equation \ref{newfitness})
$$
W(z) = \beta^{\star}{\rm e}^{-(z-m)^{2}/2\omega^{2}}, \quad\beta^{\star}=\beta\times{\rm e}^{-(z_{F}^{\star}-z_{S}^{\star})^{2}/2\left(\frac{1}{\delta} + \frac{1}{\gamma}\right)},
$$
\noindent and assuming a normal distribution of phenotypes
$$
p(z,t) = \frac{1}{\sqrt{2\pi\sigma^{2}}} {\rm e}^{-(1/2\sigma^{2})(z-\Wb{z}(t))^2},
$$
note that the numerator and denominator become
$$
A = \int z {\rm e}^{-(z-m)^{2}/2\omega^{2}}{\rm e}^{-(1/2\sigma^{2})(z-\Wb{z}(t))^2}dz\quad\text{and}
$$
\noindent
$$
B = \int {\rm e}^{-(z-m)^{2}/2\omega^{2}}{\rm e}^{-(1/2\sigma^{2})(z-\Wb{z}(t))^2}dz,
$$
\noindent respectively. Thus, using Result 1 (equation \ref{Rslt1}) with $\mu_{1} = m =  \frac{\frac{1}{\gamma}z_{F}^{\star} + \frac{1}{\delta}z_{S}^{\star}}{\frac{1}{\delta} + \frac{1}{\gamma}}$, $\sigma^{2}_{1} = \omega^{2}$, $\mu_{2} = \Wb{z}(t)$, and $\sigma^{2}_{2} = \sigma^{2}$, we get that 
$$
A = {\rm e}^{-\frac{(\mu_{1}-\mu_{2})^{2}}{2(\sigma^{2}_{1} + \sigma^{2}_{2})}}\sqrt{2\pi\nu^{2}}m^{\star}, \quad B = \Wb{W} =  {\rm e}^{-\frac{(\mu_{1}-\mu_{2})^{2}}{2(\sigma^{2}_{1} + \sigma^{2}_{2})}}\sqrt{2\pi\nu^{2}}.   
$$
\noindent Then, the average phenotype after selection under a dual component fitness function is
\begin{align*}
\Wb{z_{W}}(t) &= \frac{A}{B} = m^{\star}  = \frac{(\sigma^{2}_{2}\mu_{1} + \sigma^{2}_{1}\mu_{2})}{\sigma^{2}_{1} + \sigma^{2}_{2}}\\
&= \frac{\omega^{2}\left(\sigma^{2}(\delta z_{F}^{\star} + \gamma z_{S}^{\star}) + \Wb{z}(t)\right)}{\sigma^{2} + \omega^{2}}\\
&= \frac{\omega^{2}}{\sigma^{2}+\omega^{2}}\bigg[ \Wb{z}(t) + \sigma^{2}(\delta z_{F}^{\star} + \gamma z_{S}^{\star})\bigg],
\end{align*}
\noindent which leads to equation \ref{newaphas}.

By plugging in the last expression into the breeder's equation, the average change in the value of a phenotypic character in response to selection in one generation under the dual fitness component is computed as:
\begin{align*}\label{MyBreeders}
\Wb{z}(t+1) - \Wb{z}(t) &=(\Wb{z_{W}}(t) - \bar{z}(t))h^{2}\nonumber\\
\nonumber\\
&= \left[\frac{\omega^{2}}{\sigma^{2}+\omega^{2}}(\sigma^{2}(\delta z_{F}^{\star} + \gamma z_{S}^{\star}) + \bar{z}(t)) - \bar{z}(t)\right] h^{2}\nonumber\\
\nonumber\\
&= \frac{h^{2}\sigma^{2}}{\omega^{2} + \sigma^{2}}\left(\omega^{2}[\delta z_{F}^{\star} + \gamma z_{S}^{\star}] - \Wb{z}(t)\right) \nonumber\\
\nonumber \\
&=\frac{h^{2}\sigma^{2}}{\omega^{2} + \sigma^{2}}\left(\mathcal{P}z_{F}^{\star} + (1-\mathcal{P})z_{S}^{\star} -\Wb{z}(t)\right).
\end{align*}

\section*{Acknowledgments}

\bibliography{darin}

\end{document}